\documentclass[reprint,floatfix,aps,amsmath,amssymb,nofootinbib,twoside,prd,showkeys,superscriptaddress]{revtex4-2}
\pdfoutput=1
\usepackage{verbatim}
\usepackage{comment}
\usepackage{graphicx}
\usepackage{epsfig}
\usepackage{bm}
\usepackage{tensor}
\usepackage{algorithm}
\usepackage{placeins}

\newcommand{\diff}{{\rm d}}

\usepackage[ddmmyyyy]{datetime} %Displaying last compiled time
\usepackage{amsmath} % Package for \eqref
\usepackage[dvipsnames]{xcolor}% Package for \textcolor
\usepackage{cancel} % Package for \cancelto{0}{x}
\usepackage{graphicx}
\usepackage{comment} %SA added 14/01/2019
\usepackage{soul, ulem}
\usepackage{eqnarray}
\usepackage{multirow}
\soulregister\cite{7}
\soulregister\ref{7}
\soulregister\eqref7
\usepackage{bbding}
\usepackage{pifont}
\usepackage{wasysym}
\usepackage{amssymb}
\usepackage{textcomp}
\usepackage{
    marginnote
} %\marginpar{xyz} \marginnote{text}[3cm] \reversemarginpar
\usepackage[
    colorinlistoftodos
    ,
    textsize=tiny
]{todonotes} %\todo{text}
\usepackage{textcomp}
\usepackage{lineno}
\usepackage{hyperref}
\usepackage{cleveref}
\hypersetup{
    colorlinks=true,
    linkcolor=blue,
    citecolor=blue,
    filecolor=magenta,
    urlcolor=blue,
    urlbordercolor=red
}

\newcommand*{\aap}{Astron.~Astrophys.}

\newcommand*{\pasp}{PASP}

\newcommand{\ksM}{\text{km/s Mpc$^{-1}$}}
\newcommand{\Mpc}{{\rm Mpc}}

\newcommand{\kb}{k_{\rm B}}
\newcommand{\Rh}{R_{\rm h}}

\newcommand{\Sbh}{\mathcal{S}_{\text{BH}}}
\newcommand{\Sfour}{\mathcal{S}_{\rm 4}}
\newcommand{\Sthree}{\mathcal{S}_{\rm 3}}
\newcommand{\LCDM}{{\Lambda \rm{CDM}}}

\newcommand{\Omzero}{\Omega_{\rm m0}}

\newcommand{\Ogt}{\Omega_{\rm GT}}

\newcommand{\Hrec}{H_{\rm rec}}
\newcommand{\rd}{r_{\rm d}}
\newcommand{\Mb}{M_{\rm b}}

\newcommand{\Hzero}{H_{0}}

\newcommand{\aplus}{\alpha_{\rm +}}
\newcommand{\aminus}{\alpha_{\rm -}}

\newcommand{\DeltaBE}{\log{\mathcal{B}}}

\newcommand{\Panp}{Pan$^{+}$}

\begin{document}
    \title{ Constraints on Generalized Gravity-Thermodynamic Cosmology from DESI DR2 }

    \author{Udit K. Tyagi}
    \email{u.tyagi@unsw.edu.au} \affiliation{School of Physics, University of New South Wales, Kensington, NSW 2032, Australia}
    \affiliation{School of Physics, Indian Institute of Science Education and Research Thiruvananthapuram, Maruthamala PO, Vithura, Thiruvananthapuram 695551, Kerala, India}

    \author{Sandeep Haridasu}
    \email{sharidas@sissa.it} \affiliation{SISSA-International School for Advanced Studies, Via Bonomea 265, 34136 Trieste, Italy}
    \affiliation{IFPU, Institute for Fundamental Physics of the Universe, via Beirut 2, 34151 Trieste, Italy}
    \affiliation{INFN, Sezione di Trieste, Via Valerio 2, I-34127 Trieste, Italy}

    \author{Soumen Basak}
    \email{sbasak@iisertvm.ac.in} \affiliation{School of Physics, Indian Institute of Science Education and Research Thiruvananthapuram, Maruthamala PO, Vithura, Thiruvananthapuram 695551, Kerala, India}

\begin{abstract}
We explore the cosmological implications of generalized entropic models within the framework of Gravity–Thermodynamics (GT) approaches. These models, characterized by three or four additional free parameters, are designed to capture deviations from the standard Bekenstein–Hawking entropy and can reproduce well-known entropic formulations, including Tsallis, Rényi, Sharma–Mittal, Barrow, Kaniadakis, and Loop Quantum Gravity entropies in various analytical limits. We implement the corresponding cosmological models using a fully numerical GT approach to constrain the model parameters and to study the evolution of the dark energy equation of state as a function of the scale factor. Our Bayesian analysis, which incorporates the Pantheon+ and DESy5 supernovae data alongside the recently released DESI-DR2/DR1 Baryon Acoustic Oscillation (BAO) measurements, shows that the data favor the standard Bekenstein–Hawking entropy, leading to a $\Lambda$CDM-like late-time behavior. In this context, the three-parameter ($\mathcal{S}_3$) entropic model appears to be sufficient to capture the observed dark energy phenomenology. Furthermore, a direct comparison of the Bayesian evidence indicates that the three-parameter model is preferred over the four-parameter ($\mathcal{S}_4$) variant by a factor of $\Delta\log\mathcal{B} \sim -6$, while the GT approach as a whole is significantly disfavored relative to the $\Lambda$CDM model with at least $\Delta\log\mathcal{B} \sim -8$ ($\mathcal{S}_3$) to $\Delta\log\mathcal{B} \sim -13$ ($\mathcal{S}_4$), when using the DESy5 and DESI-DR2 datasets.
\end{abstract}
%%%%%%%%% Abstract %%%%%%%%%%

    \date{\today\ at \currenttime\ of Roma}
    \maketitle

%%%%%%%%%%%%%%%%%%%%%%%%%%%%%%%%%%%%%%
\section{Introduction}
\label{sec:Intro}
%%%%%%%%%%%%%%%%%%%%%%%%%%%%%%%%%%%%%%

The standard Bekenstein–Hawking (BH) entropy has long served as a cornerstone in our understanding of black hole thermodynamics. According to this foundational framework, a black hole's entropy is proportional to its horizon area \cite{Hawking:1975vcx, PhysRevD.7.2333, Bardeen:1973gs, Bekenstein:1994bc}. When first proposed, this result was considered anomalous, since the entropy of conventional thermodynamic systems typically scales with volume rather than surface area. Over the years, this unique non-extensive property of black holes, along with its deep connection to strongly quantum-entangled $d$-dimensional systems, has been extensively studied \cite{Tsallis_2013, Zhang:2015gda, Zhang:2017aqf, Majhi:2017zao, Chunaksorn:2025nsl, e22010017, Elizalde:2025iku}.

More recently, numerous generalizations of the BH entropy have been proposed \cite{Nojiri:2021iko, Elizalde:2025iku, Nojiri:2023saw}. These extended formulations incorporate non-extensive statistical frameworks into conventional thermodynamics. Notable examples include Tsallis entropy \cite{tsallis1988possible}, Rényi entropy \cite{Rényi}, Barrow entropy \cite{Barrow:2020tzx}, Sharma–Mittal entropy \cite{sharma1975new, sharma1977new}, Kaniadakis entropy \cite{Kaniadakis:2005zk}, and entropies inspired by Loop Quantum Gravity \cite{Czinner:2015eyk, Majhi:2017zao, Mejrhit:2019oyi, Liu:2021dvj}. For a detailed discussion, see \cite{Nojiri:2022aof, Nojiri:2023bom, Nojiri:2024zdu}. Despite their diverse formulations, these extended entropy models share universal characteristics: they reduce to the standard BH entropy under specific limiting conditions and exhibit monotonic growth with respect to the BH variable \cite{Nojiri:2021iko, Nojiri:2022dkr}. This universal behavior suggests the potential existence of a unified entropy framework that encompasses all these models.

In recent works \cite{Odintsov:2023vpj, Odintsov:2023qfj, Nojiri:2024zdu, Nojiri:2022dkr, Bolotin:2023wiw, Nojiri:2021iko}, unified frameworks have been proposed that combine various entropy models into a single formulation, known as the "Four-Parameter Entropy." This generalized entropy, governed by four free parameters, is capable of reproducing all the aforementioned entropy forms in particular limits. Moreover, the Four-Parameter Entropy can be further reduced to a Three-Parameter Entropy, which encompasses most entropy models except Kaniadakis entropy \cite{Odintsov:2023vpj}. Since its inception, this unified framework has been extensively studied for its cosmological implications \cite{Nojiri:2022dkr, Elizalde:2025iku}.

Notably, it has been demonstrated in \cite{Nojiri:2022dkr} that entropic cosmology based on these generalized entropy models offers a unified description of the universe's evolution from an early inflationary phase, characterized by a quasi–de Sitter expansion ending near 58 e-folds, to a late-time dark energy–dominated era consistent with recent Planck data \cite{Planck18_parameters}. In a separate study, \cite{Odintsov:2023vpj} examined the transition from inflation to reheating, showing that the entropic energy density drives inflation and decays smoothly into relativistic particles during reheating. The inclusion of entropic parameters facilitates a seamless evolution of the Hubble parameter, transitioning from the inflationary quasi–de Sitter phase to a power-law reheating phase characterized by a constant equation-of-state parameter.

The entropic models discussed above have been widely applied in constructing cosmological models \cite{Odintsov:2023vpj, Leon_2021, Naeem:2023tcu, Saridakis_2018, Nojiri:2022aof, Nakarachinda:2023jko, Fazlollahi:2022bgf, KordZangeneh:2023syq, Salehi:2023zqg, Nojiri:2021jxf, Nojiri:2019skr, Nojiri:2022dkr, Odintsov:2022qnn, Nojiri:2022nmu, Nojiri:2023nop, Nojiri:2023wzz, Manoharan:2024thb, Manoharan:2022qll}. These models draw inspiration from black hole thermodynamics by extending its principles to the cosmological horizon. Two primary frameworks are used to construct such models: the \textbf{Holographic Principle} \cite{Bousso:1999dw, Bousso:2002ju, Li_2004, WANG20171, Abdalla:2001as, Susskind:1994vu, Coriano:2019eif, Granda:2011wa} and the \textbf{Gravity–Thermodynamics Principle} \cite{Jacobson:1995ab, Cai:2005ra, Padmanabhan:2003gd, Padmanabhan:2009vy}. Both methods rely on drawing an analogy between the black hole event horizon and the cosmological horizon to reveal the thermodynamic properties of spacetime. However, different entropy formulations could potentially yield distinct cosmological predictions, raising the important question: which entropy, and by extension, which model, is most favored by the observational data? 

In our previous work \cite{Tyagi:2024cqp}, we demonstrated that cosmological models derived from the Holographic Principle (see also \cite{Li:2024qus,Li:2024hrv} for a recent discussion) achieve marginally better Bayesian evidence when compared to those constructed using the Gravity-Thermodynamics Principle. Following which, 
in this work we further test the Gravity-Thermodynamic approach's viability\footnote{See also \cite{Carroll:2016lku} for a discussion on the limitations of the GT approach, where it has been suggested that it cannot provide a consistent description the entropy.} utilizing the most-recent late-time data. To address this, we adopt the Gravity–Thermodynamics (GT) Principle to construct cosmological models based on the generalized Four-Parameter Entropy and its reduced Three-Parameter form. As an added advantage, the generalized entropy formalism allows one to perform model selection without having to rely on testing each model individually against the data, which is in turn one of our primary motive. 

Also, given the more recent Baryon Acoustic Oscillation (BAO) data from DESI-DR2 \cite{DESI:2025zgx} along with the DESy5 \cite{DES:2024tys} and UNION3 \cite{Rubin:2023ovl}, provides strong hints for the phantom crossing and dynamical dark energy \cite{DESI:2025fii, Gu:2025xie}\footnote{This strong hint for a dynamical nature of dark energy has given inspired several recent investigations, see for instance \cite{Adolf:2024twn,Akarsu:2025gwi,Berti:2025phi, Wolf:2025jed,Li:2025cxn,DESI:2025wyn,Colgain:2025nzf,Akrami:2025zlb,Shlivko:2025fgv,Yan:2025pgp,You:2025uon,Kessler:2025kju,Pan:2025qwy,Chaussidon:2025npr,Silva:2025hxw,Nesseris:2025lke,Shah:2025ayl,Hur:2025lqc,Paliathanasis:2025dcr,Pan:2025psn,Ishiyama:2025bbd,Brandenberger:2025hof,Ormondroyd:2025iaf}, for a non-extensive lists of recent works.}, one could comprehensively conclude if the generalized GT approach is capable of capturing something similar. We therefore, also investigate the dark energy (DE) equation of state parameter $w_{\rm DE}$ and its evolution with redshift in the context of the generalized GT approach. In this context, the Holographic approach is very well known to provide a freezing-like behavior for the dark energy equation of state, completely missing the phantom crossing \cite{Li_2004, WANG20171,Li:2024hrv,Tyagi:2024cqp,Saridakis_2018} (see also \cite{Colgain:2021beg}), which however, deserves an independent assessment. 

% This method is not only more computationally tractable but also yields results that are only marginally less robust than those obtained via the holographic approach, making it a practical and plausible choice.

The structure of this paper is as follows: \Cref{sec:Model} describes the cosmological modeling of both the Three-Parameter and Four-Parameter entropy models, \Cref{sec:data} outlines the observational datasets used, \Cref{sec:results} presents the methodology and results, and \Cref{sec:conclusions} summarizes our conclusions. Unless otherwise stated, all expressions are given in natural units ($\kb = c = \hbar = 1$).

%%%%%%%%%%%%%%%%%%%%%%%%%%%%%%%%%%%%%%
\section{Thermodynamic Gravity}
\label{sec:Model}
%%%%%%%%%%%%%%%%%%%%%%%%%%%%%%%%%%%%%%

In this section, we present an overview of the modeling of cosmological observables within the gravity–thermodynamics framework. This approach has been discussed extensively in the literature (see, e.g., \cite{Nojiri:2005pu, Nojiri:2022dkr, Tyagi:2024cqp, Leon:2021wyx, Saridakis_2020}).

%%%%%%%%%%%%%%%%%%%%%%%%%%%%%%%%%%%%%%%
\subsection{Three-Parameter Generalized Entropy Model}
\label{sec:three_para}
%%%%%%%%%%%%%%%%%%%%%%%%%%%%%%%%%%%%%%%

We begin by considering a flat FLRW universe (i.e., setting $k=0$), for which the radius of the cosmological horizon is
\begin{equation}
    \Rh = \frac{1}{\sqrt{H^{2} + \frac{k}{a^2}}} = \frac{1}{H}\,.
\end{equation}
where $H$ denotes the Hubble parameter.

The first generalized entropy we explore, which governs the dynamics in the bulk, is assumed to be \cite{Nojiri:2022aof}
\begin{equation}
    \Sthree = \frac{1}{\gamma} \Biggl[\Bigl(1 + \frac{\alpha}{\beta}\, \Sbh\Bigr)^{\beta} - 1\Biggr],
\end{equation}
where $\Sbh$ is the standard Bekenstein–Hawking entropy defined on the event horizon \cite{Bekenstein:1972tm,Bekenstein:1993dz}:
\begin{equation}
    \Sbh = \frac{A}{4\,G_{\rm N}}, \quad \text{with} \quad A = 4\pi\,\Rh^{2}\,.
\end{equation}
In these expressions, $G_{\rm N}$ is Newton's gravitational constant and $A$ represents the surface area of the horizon (or the apparent horizon in the cosmological context).

The incremental change in energy across the horizon is given by
\begin{equation}
    \label{dQ}
    \diff Q = -\diff E = -\frac{4\pi}{3}\,\Rh^{3}\,\dot{\rho}\,\diff t
    = -\frac{4\pi}{3H^{3}}\,\dot{\rho}\,\diff t\,,
\end{equation}
where $\dot{\rho}$ denotes the time derivative of the energy density. Using the Gibbons–Hawking temperature,
\[
T = \frac{H}{2\pi}\,,
\]
and the first law of thermodynamics, $\diff Q = T\,\diff S_{G}$, we obtain the relation
\begin{equation}
    -\frac{\alpha}{G_{\rm N}\,H^{2}\,\gamma}\left(1 + \frac{\pi\alpha}{G_{\rm N}\,\beta\, H^{2}}\right)^{\beta - 1}\diff H
    = -\frac{4\pi}{3H^{3}}\,\dot{\rho}\,\diff t\,.
\end{equation}

After rearranging terms and integrating both sides, we arrive at the modified Friedmann equation:
\begin{equation}
    H^{2} = \frac{8\pi\,G_{\rm N}}{3}\Bigl(\rho + \rho_{\rm GT}\Bigr) + \frac{\Lambda}{3}\,,
\end{equation}
with the additional energy density component defined as
\begin{equation}
    \begin{split}
        \rho_{\rm GT} = \frac{3}{8\pi\,G_{\rm N}}\Biggl[\,H^{2} - \frac{\pi\alpha^{2}}{(2-\beta) \,G_{\rm N}\,\beta\, \gamma}\, \Biggl(\frac{G_{\rm N} H^{2}\beta}{\pi\alpha}\Biggr)^{2-\beta}\\
        \qquad\qquad\times\, {}_{2}F_{1}\Bigl(1-\beta,\, 2-\beta,\, 3-\beta,\,-\frac{G_{\rm N}H^{2}\beta}{\pi\alpha}\Bigr)\Biggr]\,.
    \end{split}
\end{equation}

This result can be cast into the dimensionless form
\begin{equation}
    E^{2} \equiv \frac{H^{2}}{H_{0}^{2}} = \Omega_{m0}(1+z)^{3} + \Omega_{r0}(1+z)^{4} + \Omega_{\Lambda} + \Ogt(z)\,,
\end{equation}
where the dimensionless dark energy density $\Ogt(z)$ is defined as
\begin{equation}
    \label{Omegagt}
    \begin{split}
        \Ogt(z) = \frac{1}{H_{0}^{2}} \Biggl[\, H^{2} - \frac{\pi\alpha^{2}}{(2-\beta)G\,\beta\,\gamma}\, \Biggl(\frac{G\,H^{2}\beta}{\pi\alpha}\Biggr)^{2-\beta}\\[1ex]
        \qquad\qquad\times\, {}_{2}F_{1}\Bigl(1-\beta,\, 2-\beta,\, 3-\beta,\,-\frac{G\,H^{2}\beta}{\pi\alpha}\Bigr)\Biggr]\,.
    \end{split}
\end{equation}

Because solving for $H(z)$ analytically is intractable due to its implicit appearance on both sides, we instead solve numerically the set of coupled differential equations. For example, the redshift derivative of $H$ is given by
\begin{equation}
    \label{H_prime}
    H' = \frac{H_{0}^{2}}{2H}\Bigl[3\Omega_{m0}(1+z)^{2} + 4\Omega_{r0}(1+z)^{3} + \Ogt'(z)\Bigr]\,,
\end{equation}
where the prime denotes differentiation with respect to $z$. Differentiating $\Ogt(z)$ and substituting into \eqref{H_prime} yields
\begin{equation}
    \label{omega_g_p}
    \Ogt'(z) = \frac{2H\,H'}{H_{0}^{2}}\left[1 - \frac{\alpha}{\gamma}\left(\frac{G_{\rm N}\,\beta\,H^{2}}{\pi\alpha + G_{\rm N}\,\beta\,H^{2}}\right)^{1-\beta}\right]\,.
\end{equation}
Substituting the expression for $H'$ into the above relation allows us to recover $\Ogt(z)$ by solving the resulting differential equation.

Finally, the dark energy equation of state (EoS) is defined by
\begin{equation}
    w_{\rm GT} = \frac{P_{\rm GT}}{\rho_{\rm GT}}\,.
\end{equation}
Using the conservation law,
\[
\dot{\rho}_{\rm GT} + 3H\,\rho_{\rm GT}\,(1+w_{\rm GT}) = 0\,,
\]
we obtain
\begin{equation}
    w_{\rm GT} = -\frac{\dot{\rho}_{\rm GT}}{3H\,\rho_{\rm GT}} - 1\,.
\end{equation}
Upon converting the time derivative to a derivative with respect to redshift via $\frac{dz}{dt} = H(1+z)$, this expression becomes
\begin{equation}
    w_{\rm GT} = -1 + \frac{\Ogt'(z)}{\Ogt(z)}\frac{1+z}{3}\,,
\end{equation}
with $\Ogt'(z)$ representing the derivative of the dimensionless dark energy density with respect to $z$.

%%%%%%%%%%%%%%%%%%%%%%%%%%%%%%%%%%%%%%%
\subsection{Four-Parameter Generalized Entropy Model}
\label{sec:four_para}
%%%%%%%%%%%%%%%%%%%%%%%%%%%%%%%%%%%%%%%

In addition to the three-parameter extension, an extended four-parameter model has been introduced to assess, for example, primordial gravitational waves \cite{Nojiri:2024zdu}. This model is expressed as
\begin{equation}
    \Sfour = \frac{1}{\gamma} \Biggl[\Bigl(1 + \frac{\alpha_{+}}{\beta}\, \Sbh\Bigr)^{\beta} - \Bigl(1 + \frac{\alpha_{-}}{\beta}\, \Sbh\Bigr)^{-\beta}\Biggr]\,,
\end{equation}
where the parameters $\{\alpha_{+}, \alpha_{-}, \gamma, \beta\}$ are positive. Following the same formalism above, one can derive the modified Friedmann equation. In the four-parameter case, the Friedmann equation takes the form
\begin{widetext}
    \begin{align}
        \label{four_para_derivation}
        \frac{8\pi\,G_{\rm N}\,\rho}{3} + \frac{\Lambda}{3} &= \frac{G_{\rm N}H^{4}\beta}{\pi\gamma} \Biggl[\frac{1}{2+\beta}\left(\frac{G_{\rm N}H^{2}\beta}{\pi\alpha_{-}}\right)^{\beta}\,{}_{2}F_{1}\left(1+\beta,\, 2+\beta,\, 3+\beta,\,-\frac{G_{\rm N}H^{2}\beta}{\pi\alpha_{-}}\right) \nonumber\\[1ex]
        &\quad + \frac{1}{2-\beta}\left(\frac{G_{\rm N}H^{2}\beta}{\pi\alpha_{+}}\right)^{-\beta}\,{}_{2}F_{1}\left(1-\beta,\, 2-\beta,\, 3-\beta,\,-\frac{G_{\rm N}H^{2}\beta}{\pi\alpha_{+}}\right)\Biggr]\,.
    \end{align}
\end{widetext}

In \cite{Nojiri:2024zdu}, it is demonstrated that for the $\Sfour$ model to satisfy the second law of horizon entropy, the following conditions must hold:
\begin{equation}
    \label{eqn:sfourpriors}
    \frac{\alpha_{\pm}}{\beta} > \frac{G_{\rm N}H_{1}^{2}}{\pi}, \quad \gamma > 0, \quad 0 < \beta < \frac{5}{4}\,,
\end{equation}
where $H_{1}$ represents the Hubble scale at the epoch of inflation. Although $H_{1}$ is not strongly constrained by late-time data, these conditions could potentially serve as useful priors in our MCMC analysis. However, we utilize much broader priors as described in \cref{sec:data}. As before, we solve the coupled differential equations numerically to constrain the model parameters from observational data. While there exist additional modifications to the basic model in \cite{Nojiri:2024zdu}, the extensions adopted here provide sufficiently diverse phenomenology to enable meaningful comparisons with late-time cosmological data (see \cref{sec:results}). Note that an excessive number of free parameters can lead to unconstrained scenarios, so our implementation strikes a balance between flexibility and constraint ability. 

Finally, \cref{table:entropies} summarizes the parameter regimes under which the generalized $\Sthree$ and $\Sfour$ entropies reduce to standard modified entropy forms such as $S_{\rm SM}$ \cite{sharma1975new, sharma1977new}, $S_{\rm T}$ \cite{Tsallis_2013}, $S_{\rm B}$ \cite{Barrow:2020kug, Barrow:2020tzx}, $S_{\rm R}$ \cite{Renzi:2017cbg}, $S_{\rm LQG}$, and $S_{\rm K}$ \cite{Kaniadakis:2005zk}. This table, which partly reproduces TABLE I from \cite{Nojiri:2024zdu}, is included here solely for completeness.

\begin{table*}[!ht]
    \centering
    \renewcommand{\arraystretch}{1.7}
    \begin{tabular}{|c|l|l|}
        \hline
        \textbf{System}            & \textbf{Conditions on $\alpha,\,\alpha_{\pm},\,\beta,\,\gamma$}                              & \textbf{Entropy}          \\
        \hline
        \multirow{5}{*}{$\Sfour$}  & $\alpha_{-}=0,\quad \alpha_{+}=\gamma$                                                        & $S_{\rm SM}$              \\
        \cline{2-3}                & $\alpha_{+}\to\infty,\quad \alpha_{-}=0$                                                      & $S_{\rm T}, \, S_{\rm B}$  \\
        \cline{2-3}                & $\alpha_{-}=0,\quad \alpha_{+}=\gamma,\quad \beta\to 0$ (with $\alpha_{+}/\beta$ finite)         & $S_{\rm R}$               \\
        \cline{2-3}                & $\beta\to\infty,\quad \alpha_{-}=0,\quad \alpha_{+}=\gamma$                                   & $S_{\rm LQG}$             \\
        \cline{2-3}                & $\beta\to\infty,\quad \alpha_{+}=\alpha_{-}$                                                 & $S_{\rm K}$               \\
        \hline
        \multirow{4}{*}{$\Sthree$} & $\gamma=\alpha$                                                                             & $S_{\rm SM}$              \\
        \cline{2-3}                & $\alpha\to\infty$                                                                           & $S_{\rm T}, \, S_{\rm B}$  \\
        \cline{2-3}                & $\alpha,\beta\to0$ (with $\alpha/\beta$ finite)                                              & $S_{\rm R}$               \\
        \cline{2-3}                & $\beta\to\infty,\quad \gamma=\alpha$                                                        & $S_{\rm LQG}$             \\
        \hline
    \end{tabular}
    \caption{Conditions on parameters $\alpha$, $\alpha_{\pm}$, $\beta$, and $\gamma$ under which the generalized $\Sthree$ and $\Sfour$ entropies reduce to standard modified forms: $S_{\rm SM}$ \cite{sharma1975new, sharma1977new}, $S_{\rm T}$ \cite{Tsallis_2013}, $S_{\rm B}$ \cite{Barrow:2020kug, Barrow:2020tzx}, $S_{\rm R}$ \cite{Renzi:2017cbg}, $S_{\rm LQG}$, and $S_{\rm K}$ \cite{Kaniadakis:2005zk}.}
    \label{table:entropies}
\end{table*}

%%%%%%%%%%%%%%%%%%%%%%%%%%%%%%%%%%%%%%
\section{Dataset and Analysis}
\label{sec:data}
%%%%%%%%%%%%%%%%%%%%%%%%%%%%%%%%%%%%%%

In this section, we briefly describe the datasets employed to constrain the models introduced earlier.

\subsection{SNe Datasets}
We utilize the well-established Pantheon+ (\Panp) dataset, which comprises 1701 light curves of 1550 spectroscopically confirmed Type Ia supernovae. This compilation includes an extensive review of redshifts, peculiar velocities, photometric calibrations, and intrinsic-scatter models, with the supernovae observed in the redshift range $0.001 \leq z \leq 2.26$. When constraining the models using only the \Panp dataset, we apply a prior on the absolute magnitude, $\Mb = -19.253 \pm 0.029$, which is equivalent to the Cepheid calibration presented in \cite{Scolnic:2021amr}.

In addition, we also utilize the more recent DESy5 dataset, which is composed of supernovae collected during the five-year DES Supernova Program. The classification of these supernovae is achieved using a machine learning algorithm applied to their light curves across four photometric bands \cite{DES:2024haq}. The DESy5 sample consists of 1830 Type Ia supernovae spanning a redshift range of $0.02 \leq z \leq 1.13$. Although the DESy5 sample has not been scrutinized as rigorously as the \Panp dataset, our aim is to refine the constraints and compare them with recent findings on dynamical dark energy reported in \cite{DESI:2024mwx}\footnote{Note that in our joint analysis, both the \Panp and DESy5 datasets are treated as uncalibrated. However, when analyzed individually, with the \Panp dataset we utilize the $\Mb$ prior, while the DESy5 dataset accounts for absolute magnitude calibration through analytical marginalization, which precludes a direct determination of $\Hzero$ \cite{DES:2024haq}.}, where the former plays a very crucial role in providing higher significance for the evidence in favor of phantom-crossing.

\subsection{Baryon Acoustic Oscillations}
We employ the most recent BAO measurements from the DESI data release (DESI-DR2)\footnote{We also report constraints using DESI-DR1 \cite{DESI:2024mwx}, noting that much of the analysis preceding DR2 was performed with this earlier release.} \cite{DESI:2025zgx}. This dataset provides measurements of the transverse comoving distance and the Hubble rate, relative to the sound horizon ($\rd$), in six redshift bins over the range $0.1 \leq z \leq 4.2$, based on observations of over six million extragalactic objects. For a complete summary, see Table IV in \cite{DESI:2025zgx}. We use this uncalibrated BAO dataset in conjunction with the inverse distance ladder method \cite{Addison:2017fdm,Aubourg_2015,Haridasu18}, adopting priors on the sound horizon obtained from CMB data \cite{Planck18_parameters} and BBN constraints on baryon density ($\Omega_{\rm b}h^{2}$).

\subsection{Inverse Distance Ladder Priors}
We adopt priors on pre-recombination physics, which have been robustly shown to be independent of late-time cosmology \cite{Verde17}. Our chosen priors on $\{\rd, \Hrec\}$\footnote{Notice that at recombination redshift ($z_{\rm rec} \sim 1089$) the contribution of expansion rate will be dominated by physical matter density $\Hrec^2 \sim \Omega_m h^2 + \Omega_{\rm rad}h^2$ and be essentially equivalent to the background priors as implemented in \cite{Lemos:2023xhs}.} are comparable to, yet less stringent than, the CMB-based background priors used in \cite{Lemos:2023xhs} and the early universe priors implemented in \cite{DESI:2024mwx, DESI:2025fii} (see also \cite{Lynch:2025ine}). Specifically, we use the values and covariance from \cite{Haridasu:2020pms}, derived using Planck 2018 likelihoods \cite{Planck:2018vyg}. These priors play a crucial role in our inverse distance ladder analysis of the BAO data, exerting only a minor influence on late-time parameter constraints and the determination of $\Hzero$. In addition, we impose a parametric fitting formula for the sound horizon from \cite{Brieden:2022heh}:
\begin{equation}
     \rd \sim 147.05 \left( \frac{\omega_{\rm b}}{0.02236} \right)^{-0.13} \left( \frac{\omega_{\rm cb}}{0.1432} \right)^{-0.23} \left( \frac{N_{\mathrm{eff}}}{3.04} \right)^{-0.1}
    \label{eqn:aub}
\end{equation}
    % \rd = 55.154\, \frac{\exp{\left[-72.3\left(\Omega_{\nu} h^2 + 6\times10^{-4}\right)^2\right]}}{(\Omega_{\rm b}h^{2})^{0.12807}(\Omega_{m}h^{2}-\Omega_{\nu}h^{2})^{0.25351}}\, ,
%
where we assume $\omega_{\rm b} = \Omega_{\rm b}h^{2} = 0.02218\pm 0.00055$ \cite{Schoneberg:2024ifp}\footnote{Given the stringent constraint, we do not sample upon the baryon matter density to its mean value. Note also that this value is consistent with older estimates reported as $\Omega_{\rm b}h^{2}= 0.0217$ \citep{Riemer-Sorensen:2017vxj}, as we have utilized in \cite{Tyagi:2024cqp}.} and $\Omega_{\nu}h^{2}= 6.42\times10^{-4}$ and $N_{\rm eff} = 3.046$ \cite{Ade16}, which is utilized to compute $\omega_{\rm cb} = \omega_{\rm c} + \omega_{\rm b}$. We note that using an equivalent fitting function as in \cite{Aubourg_2015} does not alter our conclusions. Incorporating \cref{eqn:aub} alongside the $\rd$ prior based on the $\LCDM$-CMB constraints ensures that the early universe remains unaltered relative to the concordance cosmology, which is essential for attributing any deviations to late-time modifications in the background evolution. Finally, we implement a Gaussian likelihood function:
\begin{equation}
    -2 \ln{\mathcal{L}} \equiv \chi^{2}_{\rm Gau} = \Delta \boldsymbol{\mu}^{T}\, \Sigma^{-1}\, \Delta \boldsymbol{\mu}\, ,
\end{equation}
where $\mathcal{L}$ is the likelihood, $\Delta \boldsymbol{\mu}$ represents the residuals between the observed data and the theoretical predictions, and $\Sigma$ is the corresponding covariance matrix.

We perform a fully Bayesian joint analysis of the datasets using the \texttt{emcee} package \cite{Foreman-Mackey13}, which implements an affine-invariant ensemble sampler. MCMC samples are generated with \texttt{corner}\footnote{\href{https://corner.readthedocs.io/en/latest/}{https://corner.readthedocs.io/en/latest/}} and/or \texttt{GetDist}\footnote{\href{https://getdist.readthedocs.io/}{https://getdist.readthedocs.io/}} \cite{Lewis:2019xzd}. In \cref{tab:priors}, we list the priors employed in our Bayesian analysis. For the MCMC analysis, we work with the logarithms of the parameters $\alpha_{\pm}$ and $\gamma$ to span several orders of magnitude conveniently. The priors on $\{\alpha_{\pm}, \gamma, \beta\}$ are chosen to be broad enough to encompass the ranges reported in the literature for various entropy models.
 
{\renewcommand{\arraystretch}{1.4} 
\setlength{\tabcolsep}{6pt} 
\begin{table}[h!]
    \centering
    \caption{Priors used in the Bayesian analysis.}
    \label{tab:priors}
    \begin{tabular}{ccc}\hline
        \textbf{Parameter} & Model & \textbf{Prior} \\ \hline\hline
        $\Omzero$ & & $[0.1, 1.0]$\\
        $\Hzero$ & & $[60.0, 80.0]$\\
        $\log{\alpha_-}$, $\log{\alpha_+}$ & $\Sfour$ & $[-5.0, 2.0]$\\
        $\log{\alpha}$ & $\Sthree$ & $[-5.0, 2.0]$\\
        $\log{\gamma}$ & $\Sthree, \Sfour$ & $[-5.0, 2.0]$\\
        $\beta$ & $\Sthree, \Sfour$ & $[0.0, 2.0]$\\
        $M_{\rm b}$ & & $[-20.0, -18.0]$ \\ \hline
    \end{tabular}
\end{table}
}

We also assess the Bayesian evidence $\Delta \mathcal{B}$ \cite{Trotta:2017wnx, Trotta:2008qt} utilizing a modified version of \texttt{MCEvidence}\footnote{\href{https://github.com/yabebalFantaye/MCEvidence}{https://github.com/yabebalFantaye/MCEvidence}} \cite{Heavens:2017afc}. We follow the convention that a negative value of $\Delta \log\mathcal{B} = \log\mathcal{B}_{\rm I} - \log\mathcal{B}_{\rm Ref=\LCDM}$ implies that the model-${\rm I}$ under consideration is disfavored the reference ($\LCDM$) model. The usual practice to contrast with the Jefferys' scale \cite{jeffreys1998theory, Kass95} (see also \cite{Trotta:2008qt, Nesseris:2012cq}) is to consider $\Delta \log\mathcal{B} < 0$ as disfavored, $0 < \Delta \log\mathcal{B} < 1$ as weakly favored, $1 < \Delta \log\mathcal{B} < 2.5$ as moderately favored, and $\Delta \log\mathcal{B} > 2.5$ as strongly favored. We also note that the Bayesian evidence is sensitive to the prior choice and, thus should be interpreted with caution. 

\begin{figure*}
    \centering
    \includegraphics[scale=0.75]{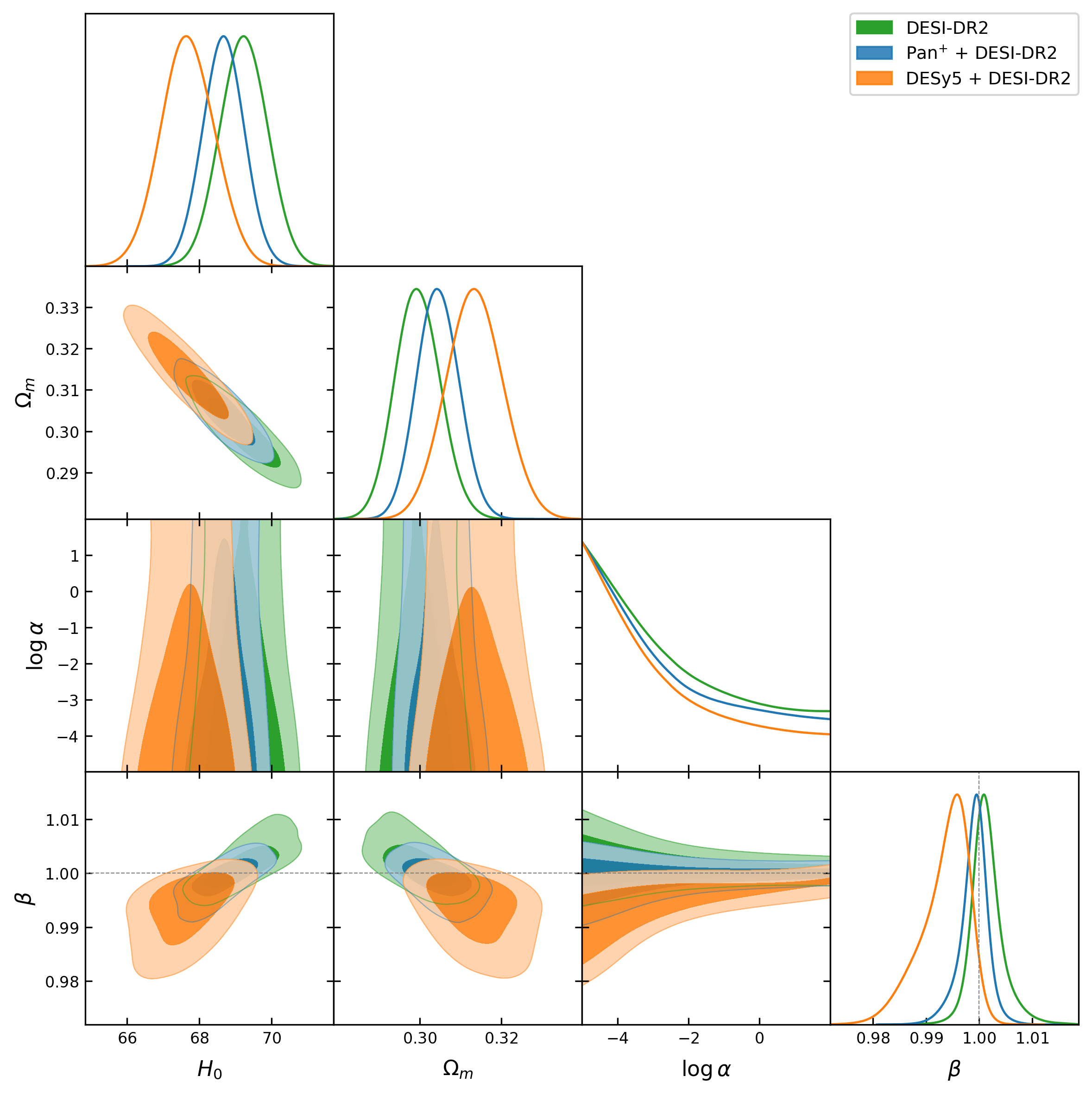}
    \caption{Contour plots for the generalized three-parameter entropy model ($\Sthree$) obtained from the DESI-DR2, DESI-DR2+Pantheon$^{+}$, and DESI-DR2+DESy5 dataset combinations. For brevity, the parameter $\gamma$—which is constrained equivalently to $\alpha$ when BAO data are included—has been omitted here (see the left panel of \cref{fig:Four_three_para_gamma} for further details).}
    \label{fig:Three_para}
\end{figure*}

\begin{figure*}
    \centering
    \includegraphics[scale=0.7]{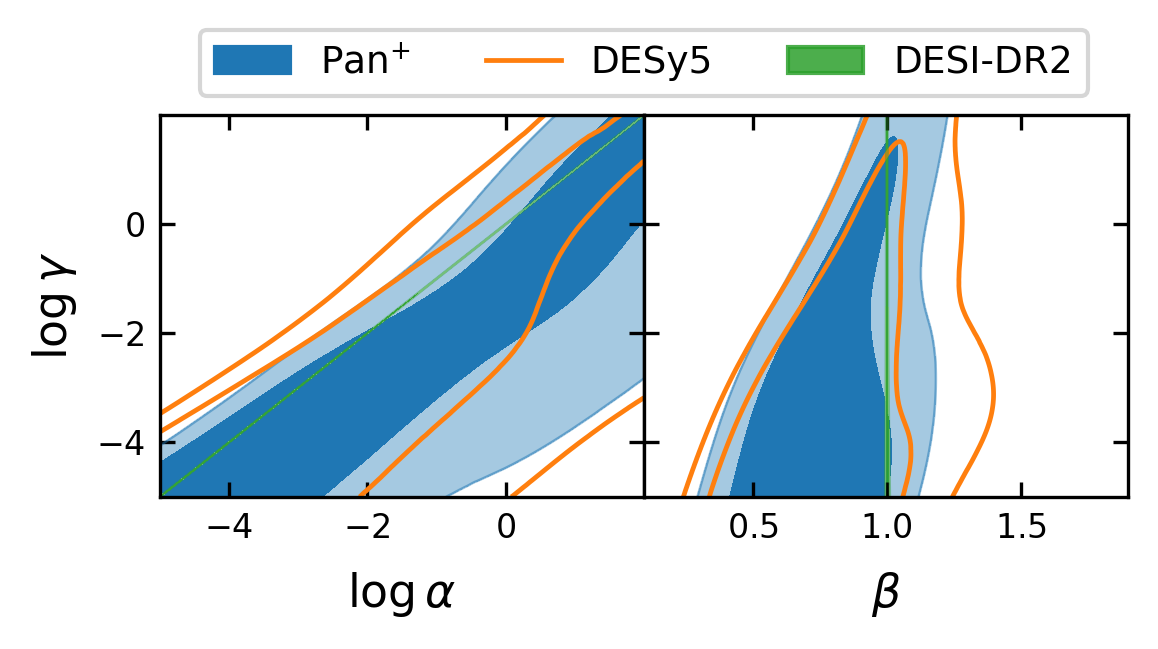}
    \includegraphics[scale=0.7]{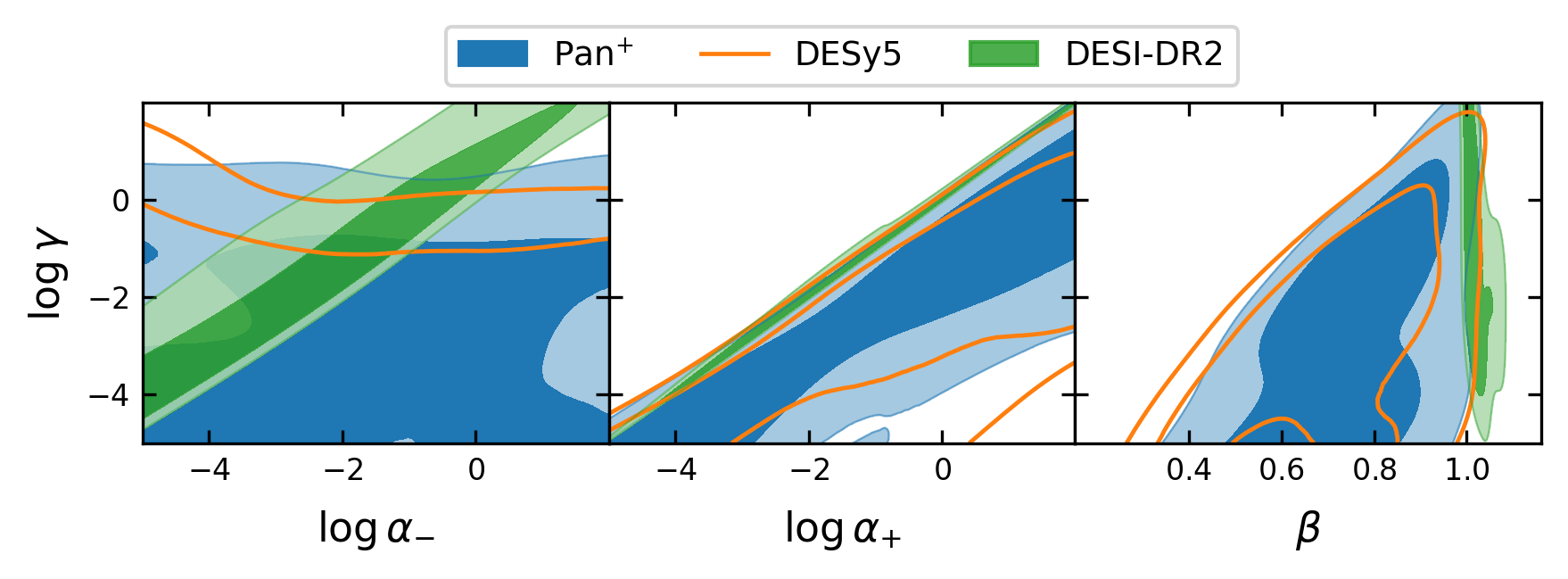}
    \caption{\textit{Left:} Contour plots for the three-parameter entropy model ($\Sthree$) displaying the parameters $\gamma$, $\alpha$, and $\beta$. \textit{Right:} Corresponding contour plots for the four-parameter entropy model ($\Sfour$), illustrating the correlation between $\gamma$, $\alpha_{\pm}$, and $\beta$. The contours are derived from the DESI-DR2, Pantheon$^{+}$, and DESy5 datasets individually.}
    \label{fig:Four_three_para_gamma}
\end{figure*}

%%%%%%%%%%%%%%%%%%%%%%%%%%%%%%%%%%%%%%
\section{Results and Discussion}
\label{sec:results}
%%%%%%%%%%%%%%%%%%%%%%%%%%%%%%%%%%%%%%

We begin by presenting the general constraints on the three-parameter ($\Sthree$) and four-parameter ($\Sfour$) entropy models (see \cref{fig:Three_para,fig:Four_para}). The corresponding constraints are summarized in \Cref{tab:table3p,tab:table4p}.

In \cref{fig:Three_para,fig:Four_three_para_gamma}, the contour plots for the $\Sthree$ entropy model are shown for various dataset combinations: DESI-DR2 alone, Pantheon$^{+}$+DESI-DR2, and DESy5+DESI-DR2. Our analysis indicates that the model parameters are reasonably well constrained overall. In particular, there is a significant correlation between the Hubble parameter $\Hzero$ and the parameter $\beta$ (and, by implication, with $\Omzero$), which tends to yield larger values of $\Hzero$ for $\beta>1.0$. However, when BAO data are included, as in the DESy5+DESI-DR2 combination, this correlation shifts toward $\beta\sim1.0$, resulting in a reduced value of $\Hzero$ of $68.42 \pm 0.77\,\ksM$. This result is in agreement with the $\Hzero$ values reported for the standard $\LCDM$ model in \cite{DESI:2025zgx} and highlights how the BAO data constrain late-time modifications that might resolve the $\Hzero$ tension, as already noted in several works \cite{Efstathiou:2020wxn,Haridasu:2020pms,Vagnozzi:2023nrq}. Furthermore, we find that the two entropy parameters, $\gamma$ and $\alpha$, exhibit a very high degree of correlation (with $\gamma / \alpha \to 1$) when BAO data are included. This strong correlation is relaxed when the constraints are derived solely from the SNe datasets, as illustrated in \cref{fig:Four_three_para_gamma}, also accompanied by the preference for $\beta < 1$. 

The four-parameter entropy encompasses the Kaniadakis entropy in addition to all the other entropies already present in $\Sthree$. As shown in \Cref{tab:table4p}, we recover similar constraints as in the $\Sthree$ model for the standard cosmological parameters. On the other hand, for the parameters describing the entropy itself, we find no constraints, while the $\aplus, \, \aminus, \, \gamma$ show a positively correlated behavior, highlighting the strong degeneracy among them. Also indicating that the flexibility\footnote{On the other hand, with all the available flexibility the GT implementation of the generalized entropies does not suffer from overfitting issues having well-defined dark energy equation of state, (see also \cref{sec:dark_energy}) behavior. } of the model is probably not constrainable by the background observables alone and might require the treatment of the perturbations as well, for instance as implemented in \cite{Lymperis:2025vup}, to break the degeneracies.  

Also, the constraints are in very good agreement with the theoretical limits on the parameters are summarized in \cref{eqn:sfourpriors}. We find that the constraint on the $\beta = 1.067^{+0.01}_{-0.02}$ is much more extended and through correlation arrives at $\Hzero \sim 72$, within $\sim 2\sigma$ confidence level, when using only the BAO DESI-DR2 data. Which is however, once again, disfavored by both the SNe datasets enforcing the index to be consistent with unity in a joint analysis. Comparing the BAO constrains from the DESI DR1 \cite{DESI:2024mwx} and DR2 \cite{DESI:2025zgx} datasets, we find that the latter provides a more stringent constraint on $\Hzero$ and $\Omzero$, as anticipated. The effect is even more pronounced for the $\Sfour$ model where DESI-DR1 constraint of $\Hzero = 70.70 \pm 1.50$ is improved to $\Hzero = 69.62 \pm 0.72$ with the more recent DR2 data. Similarly the constraints on $\beta$ also become more consistent with unity for the DR2 dataset. While at the face-value the constraints on $\beta$ are stringently constrained by the late-time BAO data, we clarify that the major improvement in the constraints is aided by the inclusion of the pre-recombination physics through the fitting formula of sound horizon in \cref{eqn:aub}, aided by the inverse distance ladder prior on $\rd$ from the CMB data. For instance, imposing only a prior on $\rd$ from the CMB data without the inclusion of the fitting formula drives the constraints to $\beta < 1$, while also providing lower values of $\Omzero < 0.3$ at $\sim 2\sigma$ confidence level. While this will be more consistent with the SNe datasets note that the SNe are unable to provide any limits on the matter density. 

As described in \cref{sec:Model}, constraints on the  generalized $\Sfour$ and $\Sthree$ model can be reduced to different entropy models as summarized in \cref{table:entropies}. Therefore, we now proceed to make the numerical comparison of the posteriors to the criteria presented in \cref{table:entropies} to evaluate the preference for the different entropy models. We find that find that both the $\Sthree$ and $\Sfour$ models capable of describing the standard $\LCDM$ model, essentially implying the standard Bekenstein-Hawking entropy, as indicated by the constraints on $\alpha \to \infty$ and $\beta$ being consistent with unity\footnote{The constraints here obtained for late-time DE phenomenology should not be contrasted immediately with the limits form early universe \cite{Nojiri:2024zdu}, where they place limits of $0.08 < \beta < 0.4$ based on inflation and reheating considerations. }. Both the generalizations $\Sthree$ and $\Sfour$ are also very-well capable of describing both the Tsallis and Barrow entropies, with the parameter $\beta = 0.9963^{+0.0054}_{-0.0029}$ and $\beta = 1.0018^{+0.0017}_{-0.0055}$, respectively, when utilizing the data combination of DESI-DR2+DESy5. The constraints on the parameter $\beta$ here can be straight away compared with the usual notations of $1+\Delta/2 \to \beta$ in the case of Barrow \cite{Barrow:2020kug, Barrow:2020tzx} and $\delta \to \beta$ in Tsallis \cite{tsallis1988possible} cases. Following which we place $68\%$ confidence level limits on the Barrow fractal index as $\Delta = -0.007^{+0.010}_{-0.006}$ and $\Delta = 0.004^{+0.003}_{-0.011}$ using $\Sthree$ and $\Sfour$ models, respectively, with DESI-DR2+DESy5 data. Note that the data combination of DESI-DR2+Pan$^{+}$ provides even tighter constraints on the parameter $\beta$ for both the models, however, we quote here slightly conservative limits using the DESI-DR2+DESy5 dataset. 

Contrasting our posteriors with the limits presented in \cref{table:entropies}, we find that that the Loop quantum gravity and the Renyi entropy are extremely disfavored as the limiting case of $\beta \to \infty$ and $\beta \to 0$, respectively, are completely ruled out by the data. Similarly, for the Kaniadakis entropy encompassed by the $\Sfour$ model alone, we find that the $\beta \to \infty$ case is disfavored even if the $\aminus = \aplus$ condition can be met within the posteriors. 

In summary, our results demonstrate that both extended entropy models are well constrained by current data, with detailed parameter correlations emerging, especially when combining BAO and SNe data. These findings provide important insights into the potential of late-time cosmological observations to test modified gravity and entropy frameworks against the standard $\LCDM$ model. However, the constraints are consistent with the $\LCDM$ model at large and can be utilized to comprehensively rule out models based on Kaniadakis, Loop quantum gravity, and Renyi entropies in the gravity-thermodynamics formalism. Following these conclusions, we now proceed to elaborate on the dark energy phenomenology.

    \begin{figure*}[htbp]
        \centering
        \includegraphics[scale=0.45]{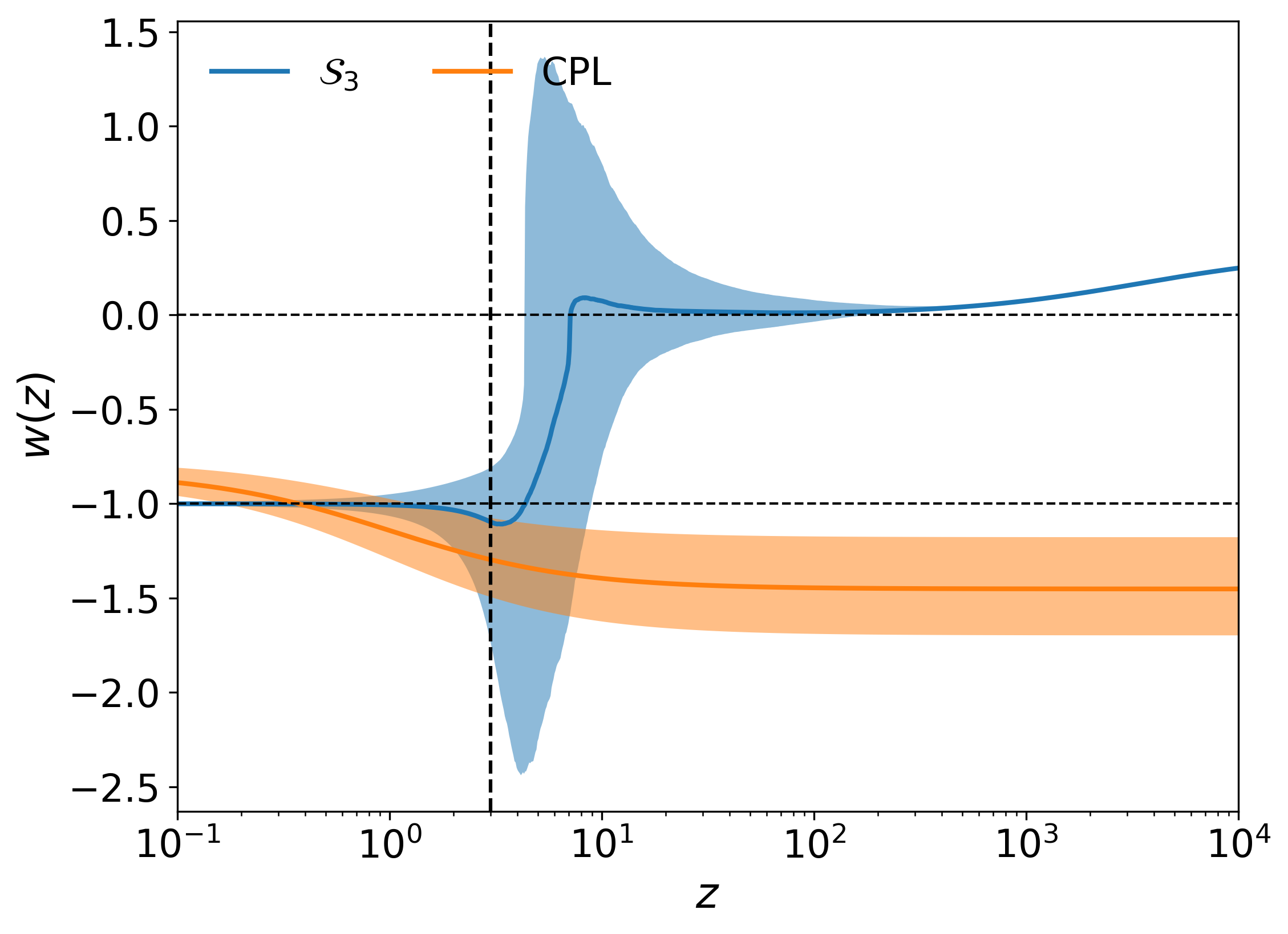}
        \includegraphics[scale=0.45]{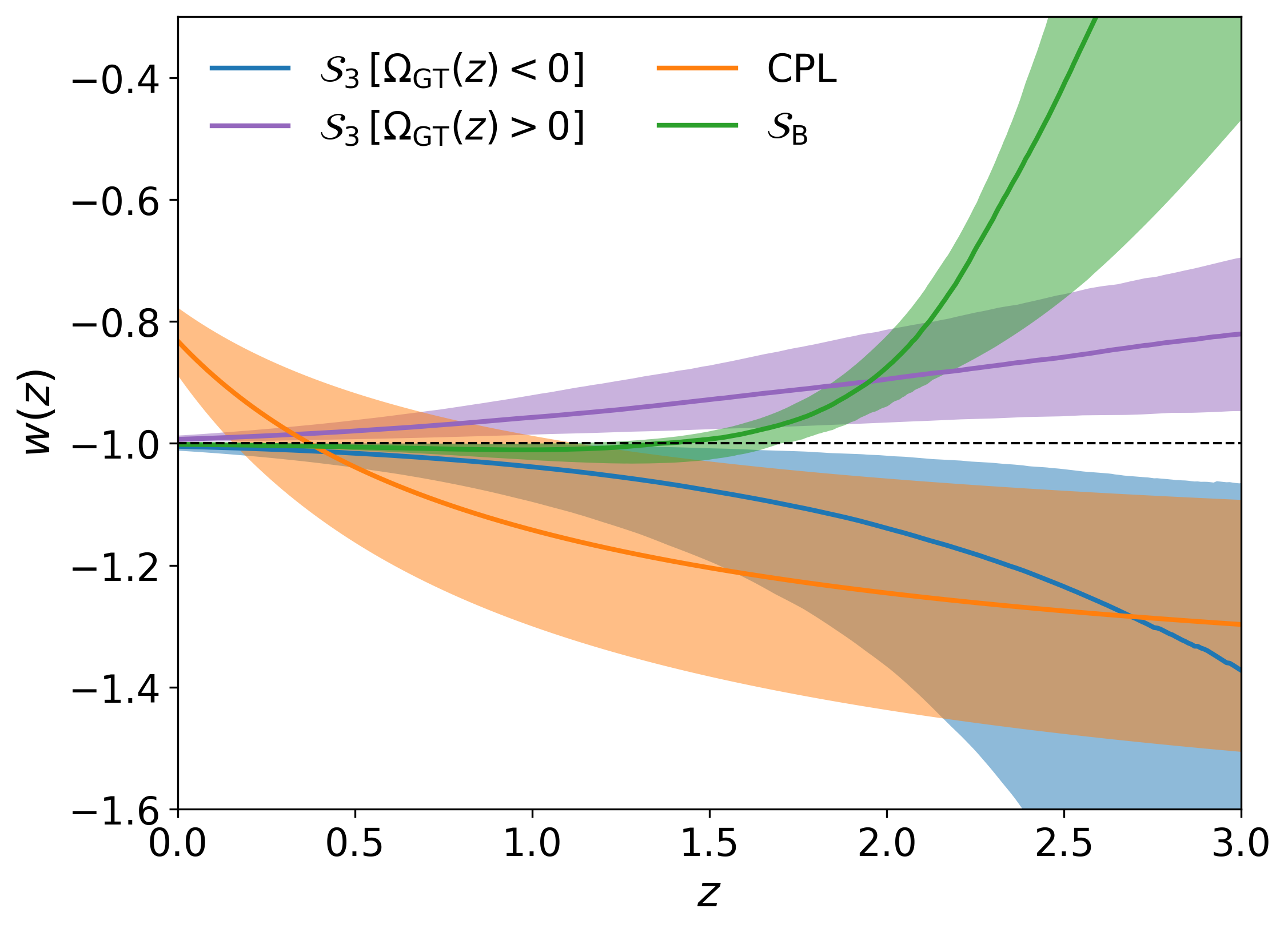}
        \caption{Evolution of the dark energy equation of state (EoS) for various scenarios within the $\Sthree$ model. \textit{Left} panel showing the extended redshift range and \textit{Right} limited to the redshift range of the late-time data. The shaded area denotes the 68\% confidence-level region obtained using the DESI-DR2 BAO and Pan+ SNe dataset. For comparison, the evolution for the Barrow entropy ($\mathcal{S}_{\rm B}$) model, as formulated in \cite{Tyagi:2024cqp}, is also shown. We also show the CPL parametrization obtained using the same dataset combination. The vertical-dashed line in the \textit{right} panel marks the tentative limit of the currently available data and coincides with the $z=3$ limit of the \textit{left} panel. }
        \label{fig:w_plots1}
    \end{figure*} 

%%%%%%%%%%%%%%%%%%%%%%%%%%%%%%%%%%%%%%
\subsection{Dark Energy Constraints}
\label{sec:dark_energy}
%%%%%%%%%%%%%%%%%%%%%%%%%%%%%%%%%%%%%%

As discussed in \cite{Denkiewicz:2023hyj}, the gravity–thermodynamics (GT) approach introduces a radiation-like correction $\LCDM$ and, consequently, a dark energy (DE) equation of state (EoS) that differs from $w = -1$. This correction results in a tracking behavior wherein the DE EoS closely follows the EoS of the dominant density component at each epoch \cite{Tyagi:2024cqp,Leon:2021wyx,Manoharan:2024thb,Manoharan:2022qll}. 

In the left panel of \cref{fig:w_plots1}, we present the reconstructed DE EoS over the redshift range $z \in [0,3]$. We split the overall posterior into two regions: $\Ogt < 0$ (blue) and $\Ogt > 0$ (purple), where the former is aided by a singularity in the DE EoS, and the latter is consistent with quintessence-like behavior in this redshift range. The shaded region indicates the $68\%$ confidence level (C.L.) regions obtained using the BAO (DESI-DR2) and SNe (\Panp) datasets. The DE EoS exhibits a transition from a matter-like state to a cosmological constant-like state, with the possibility of both quintessence and phantom behavior. Both the phantom and the quintessence branches converge to matter-like EoS ($w \to 0$) deep inside the matter-dominated regime before transitioning to the radiation-like ($w \to 1/3$) EoS. For comparison we also show the Barrow entropy model, using the DESI-DR2+\Panp datasets, as formulated in \cite{Tyagi:2024cqp,Leon:2021wyx}, which does not exhibit this singular behavior, but rather smoothly transitions from $w \to -1$ to $w \to 0$ in the redshift interval $2 \lesssim z \lesssim 4$. Also, it is mildly phantom-like at $z \lesssim 1.5$. The joint reconstruction of EoS using the complete posteriors clearly provides a cosmological constant like behavior. While Barrow entropy is clearly within the allowed parameter space of the $\Sthree$ posteriors, the marginalizing effects on the additional free parameters of the latter extended model provide different reconstructed DE EoS. 

We find that the phenomenon for $\Ogt < 0$ is subtly driven by the condition $\beta < 1$ in the generalized entropy models. Such behavior was also noticed in our earlier analysis in \cite{Tyagi:2024cqp} (see FIG 4. therein) for the Tsallis entropy formulations. In turn, as we have elaborated earlier the $\beta < 1$ is mostly driven by the SNe datasets. This is also consistent with the claims for the detection of the phantom crossing and the dynamic nature of DE, which are strongly driven by the inclusion of the SNe datasets. 

To compare with the standard analysis for dynamical dark energy in \cite{DESI:2025fii,DESI:2025zgx}, we also show the EoS for the CPL parametrization \cite{Chevallier:2000qy,Linder:2002et}\footnote{In the CPL model the DE EoS is parametrized as $w(a) = w_0 +w_{\rm a} (1-a)$.} in \cref{fig:w_plots1}. The standard CPL parametrization predicts a phantom crossing around $z \sim 0.6$, which is not supported by the generalized gravity-thermodynamics models. In the right panel of \cref{fig:w_plots1}, we show the reconstruction of the DE EoS in both the GT formalism and through CPL parametrization extended to the high redshifts. As one can notice the GT based $w(z)$ reconstructed the entire parameter space, shows a singularity-like behavior around $z\sim 7$ before tending to matter-like $w \to 0$ and then slowly evolving to radiation-like $w\to 1/3$. It is also interesting to note that this transition happens earlier in the generalized $\Sthree$ entropy model than in the Barrow entropy model. While the GT formalism does not provide the phantom crossing, as expected from the CPL parametrization of $w(z)$, it indeed tends to phantom values towards the farthest range of the presently available data $z\to 2.5$. 

It is also worth mentioning that, if a prior of $\Ogt(z) >0$ is imposed over the entire redsfhit range, it is possible to obtain tentative upper limits on the parameters of the $\Sthree$ model,  $\alpha, \gamma < 1.0$ at $2 \sigma$ confidence level. While this could be physically motivated before avoid energy densities and subsequently non-phantom DE EoS, we remain with full parameter space to make an equivalent comparison with CPL parametrization-based $w(z)$. These limits then, in turn, will be crucial for the Barrow and Tsallis entropies, which have $\alpha \to \infty$, $\aplus \to \infty$ as the limiting case within the $\Sthree$ and $\Sfour$ models.

\subsection{Model-selection}
\label{sec:model_selection}
Finally, we compute the Bayesian evidence for the models $\Sthree$ and $\Sfour$ against the $\LCDM$ model. We find that the extended parameter space of the $\Sthree$ and $\Sfour$ models is extremely disfavored over the standard $\LCDM$ model.
We find a $\DeltaBE \sim -7.80$ and $\DeltaBE \sim -7.79$ for the model $\Sthree$ model when using the DESI-DR2+\Panp and DESI-DR2+DESy5 dataset combinations, respectively. Similarly, we find $\DeltaBE \sim -13.77$ and $\DeltaBE \sim -12.64$ for the $\Sfour$ model when using the DESI-DR2+\Panp and DESI-DR2+DESy5 dataset combinations, respectively. While the $\Sfour$ model is strongly disfavored, we find the $\Sthree$ model is disfavored only at a similar level as the Barrow and Tsallis entropy models assessed in \cite{Tyagi:2024cqp}. 

In all, there is a clear indication of the strong preference against the GT approach with $\Sfour$ entropy, as the large parameter space of the extended entropy models does not yield any preference over the standard $\LCDM$ model. This also implies that there is no immediate need to extend the parameter space of entropy models in the GT approach through further generalized entropy formulations, such as the five \cite{Odintsov:2023vpj} and six parameter \cite{Nojiri:2024zdu,Nojiri:2023wzz} models proposed through a similar approach. Instead, it could be of utmost interest to explore entropy models in the context of providing a varied dark energy phenomenology at late times $w(z\to0) \neq -1$, keeping to the tracking-like\footnote{Along side the discussions limited to the GT approach, we find that a recent implementation of the gravity-glitch model in \cite{Wen:2023wes,Wen:2024orc}, provides equivalent DE EoS behavior as the current models. } behavior of the GT approach at early times.

%%%%%%%%%%%%%%%%%%%%%%%%%%%%%%%%%%%%%%
\section{Conclusions}
\label{sec:conclusions}
%%%%%%%%%%%%%%%%%%%%%%%%%%%%%%%%%%%%%%

In this work, we have applied a gravity-thermodynamics (GT) approach to implement a recently proposed generalized entropy functional form that can capture a wide range of dark energy (DE) phenomenology. We performed a comprehensive numerical analysis to constrain the model parameters using the latest BAO (DESI) and SNe (Pantheon+ and DESy5) datasets, with the inverse distance ladder priors properly incorporated. Our principal findings can be summarized as follows:

\begin{enumerate}
    \item \textbf{Consistency with $\Lambda$CDM:}  
        The three- and four-parameter extensions of the entropy models yield constraints that are in agreement with the standard $\Lambda$CDM scenario, albeit with mild deviations capable of producing a dynamical DE equation of state.
    
    \item \textbf{Reduction to Standard Entropy:}  
        Our posterior analysis demonstrates that the generalized entropy models effectively reduce to the standard Bekenstein-Hawking entropy, showing mild preference for Sharma-Mittal entropy and little to no preference for alternative modified entropy forms such as those proposed by Barrow, Tsallis, Rényi, and Kaniadakis.
    
    \item \textbf{Enhanced Dark Energy Phenomenology:}  
        Despite converging to the standard Bekenstein-Hawking entropy ($S_\text{BH}$), the extended parameter space of the $\Sthree$ and $\Sfour$ models provides a considerable range of dark energy dynamics, as illustrated in \cref{fig:w_plots1}.
    
    \item \textbf{Bayesian Evidence:}  
        We computed the Bayesian evidence for the $\Sthree$ and $\Sfour$ models. For the DESI-DR2+\Panp dataset combination, the extended parameter space does not offer any statistical advantage over the $\Lambda$CDM model, being disfavored by at least $\Delta \log B \sim -8$ for both models. Similarly, with DESy5+DESI-DR2, we find $\Delta \log B \sim -8$ and $\Delta \log B \sim -13$ for $\Sthree$ and $\Sfour$, respectively. These findings are consistent with those reported in \cite{Tyagi:2024cqp} for Barrow and Tsallis entropy models. Overall, the Bayesian analysis indicates that the large parameter space of the extended entropy models does not result in any advantage over the standard $\Lambda$CDM model. The evidence suggests a marked preference against the GT approach.
\end{enumerate}

While the extended parameter space of the $\Sthree$ and $\Sfour$ models yields a rich dark energy phenomenology, it does not lead to a statistical preference over the standard $\Lambda$CDM. Also the gravity-thermodynamics framework is not capable of mimicking the phenomenology of phantom crossing. It is of minimal interest to extend the analysis to the holographic approach using similar extended entropy formalism. Notably, given that the holographic approach often results in a near freezing–quintessence-like behavior \cite{Tyagi:2024cqp, Saridakis_2020, Nakarachinda:2023jko, Wang:2016och, Colgain:2021beg}, and in light of recent indications of a phantom crossing in the DE equation of state \cite{DESI:2024mwx, DESI:2025zgx, DESI:2025fii, Gu:2025xie}, exploring such alternatives remains an intriguing direction to investigate, however requiring necessary modifications from the current approach. In all, at a face value, we conclude that the current DESI-DR2 and DESyr5 data could be pointing away from the gravity-thermodynamics formalism, as investigated here.

    \acknowledgments
    SH is supported by the INFN INDARK grant and acknowledges support from the
    COSMOS project of the Italian Space Agency (cosmosnet.it). UKT is grateful to the IISER, Thiruvananthapuram, for hosting him during Apr'-Sep'2024 after the completion of his masters thesis, in which period the major part of the work, including the modeling and analysis with DESI-DR1 was performed.

    \appendix 
    \section{Tables of constraints and posterior contours}
    \label{sec:appendix}
\begin{figure*}
        \includegraphics[scale=0.7]{
            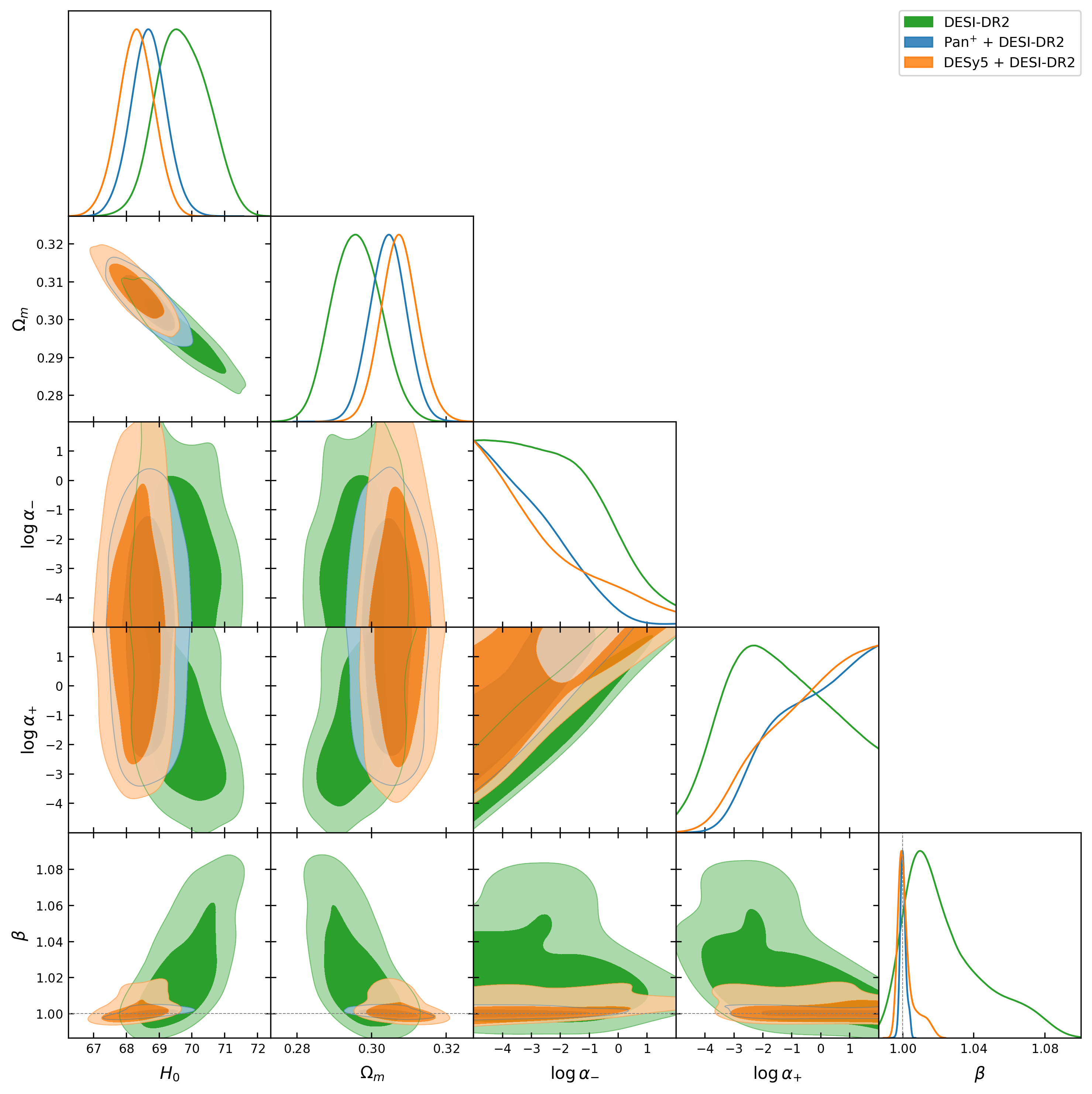
        }
        \caption{Same as \cref{fig:Three_para}, for the four parameter extended model
        ($\Sfour$). Similar to the case of $\Sthree$ entropy, here we exclude the
        $\gamma$ parameter which is highly correlated $\alpha_{+}$ when using the
        BAO datasets (see right panel of \cref{fig:Four_three_para_gamma}).}
        \label{fig:Four_para}
    \end{figure*}

    For brevity in the main text we show the table of constraints for the $\Sthree$ and $\Sfour$ in the appendix here. The contour plots of the $\Sfour$ model are shown here in \cref{fig:Four_para}.
    {\renewcommand{\arraystretch}{1.8} \setlength{\tabcolsep}{6pt}

    \begin{table*}[] \caption{Constraints (68\% C.L.) on the model parameters of the three parameter entropic model, $\Sthree$. The parameter $H_{0}$, $\rd$ are presented in the units of $\ksM$, $\Mpc$, respectively. In the runs, including the BAO data, a prior on $r_{d}= 147.09\pm 0.26 \, [{\rm Mpc}]$ is imposed. } \label{tab:table3p}
    % \centering
    \begin{tabular}{cccccccccc}\hline \textbf{Dataset} & \textbf{$\Omega_{\rm m}$} & \textbf{$H_{0}$} & \textbf{$\log{\alpha}$} & \textbf{$\beta$} & \textbf{$\log{\gamma}$}\\ 
        \hline
        \hline
    Pan$^{+}$ & $-$ & $73.30_{-1.00}^{+1.00}$ & $>-2.45$ & $0.82_{-0.19}^{+0.21}$ & $<-1.65$ \\

    DESy5 & $-$ & $-$ & $-$ & $0.83_{-0.24}^{+0.29}$ & $<-1.56$ \\

    DESI-DR1 & $0.296_{-0.0091}^{+0.0079}$ & $69.60_{-1.00}^{+1.00}$ & $<-1.47$ & $1.0_{-0.0042}^{+0.0037}$ & $<-1.47$ \\

    DESI-DR2 & $0.2995_{-0.0055}^{+0.0055}$ & $69.23_{-0.65}^{+0.65}$ & $<-1.54$ & $1.0015_{-0.0030}^{+0.0022}$ & $<-1.54$ \\

    Pan$^{+}$+DESI-DR1 & $0.306_{-0.0069}^{+0.0069}$ & $68.42_{-0.77}^{+0.77}$ & $<-1.88$ & $0.997_{-0.0027}^{+0.0042}$ & $<-1.88$ \\

    DESy5 + DESI-DR1 & $0.313_{-0.007}^{+0.007}$ & $67.73_{-0.72}^{+0.72}$ & $<-2.00$ & $0.993_{-0.0028}^{+0.0051}$ & $<-2.00$ \\

    Pan$^{+}$+DESI-DR2 & $0.3045_{-0.0051}^{+0.0051}$ & $68.67_{-0.57}^{+0.57}$ & $<-1.65$ & $0.9991_{-0.0020}^{+0.0027}$ & $<-1.66$ \\

    DESy5 + DESI-DR2 & $0.3135_{-0.0069}^{+0.0069}$ & $67.68_{-0.74}^{+0.74}$ & $<-2.08$ & $0.9936_{-0.0029}^{+0.0054}$ & $<-2.09$ \\ \hline\end{tabular}\end{table*} }

    {\renewcommand{\arraystretch}{1.8} \setlength{\tabcolsep}{6pt} 
    \begin{table*}[] \caption{Same as \cref{tab:table3p}, but for the four parameter $\Sfour$ model. The analysis is performed identical to the case of $\Sthree$.} \label{tab:table4p} \centering \begin{tabular}{cccccccc}\hline \textbf{Dataset} & \textbf{$\Omega_{\rm m}$} & \textbf{$H_{0}$} & \textbf{$\log{\alpha_{-}}$} &\textbf{$\log{\alpha_{+}}$} & \textbf{$\beta$} & \textbf{$\log{\gamma}$}\\ 
        \hline
        \hline

    Pan$^{+}$ & $-$ & $73.25_{-0.96}^{+0.96}$ & $-$ & $>-2.46$ & $0.87_{-0.24}^{+0.21}$ & $<-1.23$ \\

    DESy5 & $-$ & $-$ & $-$ & $>-1.91$ & $0.80_{-0.25}^{+0.25}$ & $<-1.48$ \\

    DESI-DR1 & $0.287_{-0.012}^{+0.012}$ & $70.70_{-1.50}^{+1.50}$ & $-1.7_{-1.8}^{+2.0}$ & $-1.3_{-1.7}^{+2.1}$ & $1.067_{-0.081}^{+0.042}$ & $-0.99_{-1.40}^{+2.20}$ \\

    DESI-DR2 & $0.2968_{-0.0061}^{+0.0061}$ & $69.62_{-0.72}^{+0.72}$ & $<-1.39$ & $-1.3_{-1.8}^{+2.0}$ & $1.0178_{-0.020}^{+0.0098}$ & $-1.2_{-2.0}^{+1.8}$ \\

    Pan$^{+}$+DESI-DR1 & $0.3047_{-0.0072}^{+0.0072}$ & $68.60_{-0.78}^{+0.78}$ & $<-1.44$ & $-1.5_{-2.0}^{+1.6}$ & $1.021_{-0.029}^{+0.011}$ & $-1.40_{-2.00}^{+1.70}$ \\

    DESy5 + DESI-DR1 & $0.3105_{-0.0077}^{+0.0077}$ & $68.02_{-0.94}^{+0.81}$ & $<-1.97$ & $>-1.52$ & $1.0141_{-0.032}^{+0.0055}$ & $>-1.46$ \\

    Pan$^{+}$+DESI-DR2 & $0.3045_{-0.0048}^{+0.0048}$ & $68.68_{-0.54}^{+0.54}$ & $<-2.59$ & $>-0.914$ & $1.0004_{-0.0018}^{+0.0012}$ & $> -0.913$ \\

    DESy5 + DESI-DR2 & $0.3075_{-0.0049}^{+0.0049}$ & $68.31_{-0.54}^{+0.54}$ & $<-2.20$ & $>-0.980$ & $1.0018_{-0.0055}^{+0.0017}$ & $>-0.969$ \\ \hline\end{tabular}\end{table*} }

    \bibliographystyle{apsrev4-1}
    \bibliography{references}

\end{document}